\documentclass[conference]{IEEEtran}
\usepackage{multirow}
\usepackage{graphicx}
\usepackage{url}
\usepackage[usenames,dvipsnames,svgnames,table]{xcolor}
\begin{document}

\title{Exploring the Mysteries of System-Level Test}

\author{\IEEEauthorblockN{Ilia Polian$^1$,
Jens Anders$^2$,
Steffen Becker$^3$,
Paolo Bernardi$^4$,
Krishnendu Chakrabarty$^5$,
Nourhan ElHamawy$^{1,2}$\\
Matthias Sauer$^6$,
Adit Singh$^7$,
Matteo Sonza Reorda$^4$ and
Stefan Wagner$^3$}
\\[0.0cm]
\IEEEauthorblockA{$^1$Institute of Computer Engineering and Computer Architecture, University of Stuttgart, Germany}
\IEEEauthorblockA{$^2$Institute of Smart Sensors, University of Stuttgart, Germany}
\IEEEauthorblockA{$^3$Institute of Software Engineering, University of Stuttgart, Germany}
\IEEEauthorblockA{$^4$Politecnico di Torino, Department of Control and Computer Engineering, Torino, Italy}
\IEEEauthorblockA{$^5$Department of Electrical and Computer Engineering, Duke University, USA}
\IEEEauthorblockA{$^6$Advantest Europe, Boeblingen, Germany}
\IEEEauthorblockA{$^7$Department of Electrical and Computer Engineering, Auburn University, USA}
}

%\IEEEoverridecommandlockouts
%\IEEEpubid{\makebox[\columnwidth]{ \hfill} \hspace{\columnsep}\makebox[\columnwidth]{ }}

\maketitle
%\IEEEpubidadjcol

\footnotetext{\copyright 2020 IEEE.  Personal use of this material is permitted.  Permission from IEEE must be obtained for all other uses, in any current or future media, including reprinting/republishing this material for advertising or promotional purposes, creating new collective works, for resale or redistribution to servers or lists, or reuse of any copyrighted component of this work in other works.}

\begin{abstract}
System-level test, or SLT, is an increasingly important process step in today's integrated circuit testing flows. Broadly speaking, SLT aims at executing functional workloads in operational modes. In this paper, we consolidate available knowledge about what SLT is precisely and why it is used despite its considerable costs and complexities. We discuss the types or failures covered by SLT, and outline approaches to quality assessment, test generation and root-cause diagnosis in the context of SLT. Observing that the theoretical understanding for all these questions has not yet reached the level of maturity of the more conventional structural and functional test methods, we outline new and promising directions for methodical developments leveraging on recent findings from software engineering. %We also discuss opportunities for SLT brought by the growing adoption of self-aware systems.
\end{abstract}

\vspace{-0.09cm}


\renewcommand{\baselinestretch}{0.94}
\normalsize


% \fi
\IEEEpeerreviewmaketitle

\vspace{-0.1cm}
\section{Introduction}
\vspace{-0.05cm}

System-Level Test (SLT) has emerged as an important additional test insertion in today’s semiconductor lifecycle \cite{Che:18}. It is run by the circuit manufacturer in the final stage of production or by the buyer of the circuit, e.g., an automotive Tier-1 supplier who will integrate the circuit into a product, as part of incoming quality control. SLT can also be used during the post-silicon characterization phase where a circuit's extra-functional properties are measured on a population of several hundreds or thousands ``first-silicon'' circuits.

Conventional structural and functional test methods are based on established theoretical concepts, such as fault models, detection and detectability concepts, coverages. A plethora of algorithms have been invented (and tools implementing these algorithms developed) in the last decades. SLT lacks much of this fundamental understanding; in fact, even the very term ``system-level test'' is being used in rather different meanings. This paper aims at making first steps towards laying solid theoretical foundations for SLT. Specifically, it discusses the following questions:
\begin{itemize}
\item What precisely is SLT that is being used in semiconductor testing? How does it differ from the traditional structural and functional test approaches?
\item What are possible reasons for \emph{SLT-unique fails}, i.e., failures observed during SLT in circuits that passed structural and functional tests during earlier test insertions?
\item How to determine the root cause of a failure during SLT, in absence of established diagnostic methods?
\item How can knowledge from the software engineering domain, e.g., on coverage definitions or on stress test generation, be 
\end{itemize}

In the remainder of the paper, we describe our current knowledge with respect to these questions, touching on related scientific disciplines where necessary. Not all questions have a known answer, and we see it as our objective to capture and discuss the currently discussed explanations or hypotheses, even if they are controversial or contradictory.

\section{System-Level Test}
\label{sec:slt}

\subsection{What is SLT?}

The term ``system-level test'' (SLT) can stand for different types of testing. In the context of integrated circuits (ICs), the following three meanings are predominant:
\begin{enumerate}
\item Test of a whole system (e.g., a smartphone or an automotive electronic control unit), focusing on interactions between its components: ICs, sensors, mechanical parts, and the like.
\item Incoming quality control of ICs by a system integrator, to sort out defective ICs and to uncover systematic quality problems of a supplier. The ICs under test are put on a board that imitates the full-system setup and applies to the IC a workload that mimics real-life operation.
\item Outgoing quality control by the IC manufacturer to prevent defective ICs from delivery and to reinforce its own quality control. The procedure is similar to 2), except that the IC manufacturer has less knowledge about the full-system setup but more knowledge about the manufactured IC.
\end{enumerate}

\begin{figure}
\includegraphics[width=0.99\columnwidth]{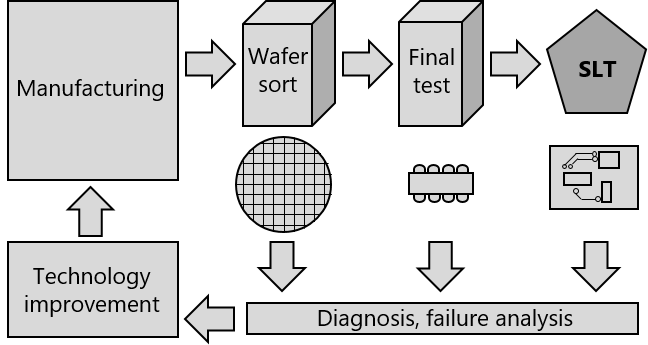}
\caption{SLT within the quality-assurance flow}
\label{fig:flow}
\end{figure}

In this paper, we focus on scenario 3), even though most findings are directly applicable to scenario 2). We do not consider scenario 1), where a failing test can point to a defect in one of the system's ICs, to a defect in a different component, or to an erroneous integration. In the SLT scenarios considered here, the system around the IC is assumed to work, and the purpose of testing is to determine whether the IC is defective or not. 
Fig.~\ref{fig:flow} visualizes the role of SLT \emph{within the IC manufacturer's quality assurance}, i.e., before the circuit has been shipped to a customer.

While there is no precise definition of ``system-level test'', it usually refers to applying to the device under test (usually, a complex system-on-chip IC) workloads that originate 
from its intended usage. A popular SLT example is booting an operating system and running several software applications known to stress the system; if the system does not behave as expected (e.g., it crashes), 
SLT has found a failure. This implies that there is currently no test specification and generation process for SLT: the workload comes from an application scenario. This may change in the future; for example, Chen \cite{Che:18} proposes to use automatically generated design-validation scenarios based on the \emph{Portable Test and Stimulus Standard} (PSS) flow.

\subsection{SLT-induced Costs and Complexities}

SLT's benefits are offset by its costs. As it is obvious from Fig.~\ref{fig:flow}, SLT is an additional test insertion that also requires special hardware. The device under test usually needs to be mounted on an evaluation board which includes memories, peripherals and interfaces necessary to run the intended workloads. It is often impossible to perform SLT on regular test equipment that does not include such features, but specialized SLT testers are available. 
Companies are currently investing in deploying SLT-oriented tester architectures that can be re-used (at least partly) over different products. In this scenario, the more diversified a company's products are, the more they profit from the re-usability of dedicated SLT testers \cite{ATS:5034}.

Another SLT cost factor is its very long test application time in the range of several minutes, a multiple of prior test insertions \cite{Che:18,Sin:19,BRR+:20sh}. This has triggered interest in \emph{adaptive test} methods, where SLT is applied only for a subset of circuits determined during earlier test insertions \cite{LTGW:18sh}. For example, Singh \cite{Sin:19} (motivated by industrial data reported in \cite{HHM+:20sh}) proposes to assign the circuits to bins based on their timing variability observed during pre-SLT test insertions. Every bin corresponds to a certain expected test quality, quantified by the number of defective circuits that will go undetected if no SLT is applied (\emph{defective parts per million} or DPPM). The expected DPPM contribution for each bin is established by a predictive model created by machine-learning from previous experiences. SLT is skipped for circuits from ``non-critical'' bins associated with (predicted) DPPM contribution below the desired DPPM target. When considering application areas where Burn-In test is required (e.g., automotive), cost reduction can be achieved by combining Burn-In test and SLT. In this case, the tester infrastructure developed for Burn-In (characterized by high parallelism) can be adapted to account for SLT requirements as well \cite{ABC+:19sh}.

\begin{table*}
\caption{SLT compared with structural and functional test}
\label{tab:slt}
\centering
\begin{tabular}{@{}|@{\hspace{0.2cm}}p{0.15\textwidth}@{\hspace{0.19cm}}|@{\hspace{0.2cm}}p{0.2\textwidth}@{\hspace{0.19cm}}|@{\hspace{0.2cm}}p{0.31\textwidth}@{\hspace{0.19cm}}|@{\hspace{0.2cm}}p{0.25\textwidth}@{\hspace{0.19cm}}|@{}}
\hline
\multicolumn{1}{@{}|@{\hspace{0.19cm}}l|@{\hspace{0.2cm}}}{\bf Aspect} & \multicolumn{1}{@{\hspace{0.2cm}}c|@{\hspace{0.2cm}}}{\bf Structural Test} & \multicolumn{1}{@{\hspace{0.2cm}}c|@{\hspace{0.2cm}}}{\bf Functional Test} & \multicolumn{1}{c|}{\bf System-Level Test} \\\hline
Level of abstraction & Gate level, sometimes incorporating additional information from layout or RTL & Instruction set architecture & None explicitly considered \\\hline
Main stimuli format & 0/1 values at circuit’s inputs / scan flops & Assembly programs  & Application or operating system code \\\hline
Test application & ATE; BIST & ATE; Software-based self-test (from cache) & Evaluation board; SLT tester \\\hline
Test generation & Fault-oriented ATPG; manual & Manually created test programs; automated techniques & Reuse of verification stimuli; applications; operating system \\\hline
Quality metrics & Fault coverage & Instruction coverage; fault coverage; coverage used in validation domain & Representative application (no explicit metric) \\\hline
How assessed? & Fault simulation; sometimes electrical simulation (Spice) of selected patterns & Instruction set level simulation, sometimes fault simulation & No systematic approach available \\\hline
What defects covered? & Gross and marginal defects represented by considered fault models and beyond & Defects not targeted by structural tests (e.g., small memories without BIST); complex defects missed by structural tests; variability; defects triggered by complex interactions within processor; a major advantage of functional test compared to structural test is the fact that it is performed at speed & Defects missed by structural and functional tests; asynchronous or analog interfaces; clock-gating logic; clock domain boundaries; unknowns (Xes); timing-related failures in uncore logic \\
\hline
\end{tabular}
\end{table*}

Fine-grained adaptive testing based on quality prediction can also be used to select a subset of chips that must undergo SLT for high-volume production~\cite{Liu:ITC2019, Liu:TODAES2020}. This strategy includes two key steps. In the first step, parametric test results from an early test insertion are used to train a machine-learning model, which can predict the quality of each chip. A random-forest model is used for quality prediction; the parametric test results of the previous test insertion are used as independent variables, and binary pass/fail results of the
current test insertion are used as dependent variables. In the second step, based on the predicted quality, chips are partitioned into two groups using k-means clustering. Test selection is performed for each group individually. SLT can be limited to chips that are predicted to be of low quality. It is shown in~\cite{Liu:ITC2019, Liu:TODAES2020} using data from three lots, including 71 wafers and 230,000 dies, that fine-grained adaptive testing reduces test cost by up to 7\% for a lot, and by as much as 90\% for low-quality chips. Moreover, experimental results also show a strong correlation between the predicted quality and marginality of the test outcomes. Therefore, the quality-prediction model can be further used to predict the occurrence of early-life failures.

\subsection{SLT vs.~Conventional Testing}

It helps the understanding of SLT to contrast its properties with conventional structural and functional test approaches. Table \ref{tab:slt} summarizes the discussion below and in the subsequent sections.

The main difference between SLT and \emph{structural test} is that the latter is strongly based on the notion of a fault according to a \emph{fault model}. Although fault models defined on various levels of abstraction have been introduced in the past \cite{NGB:99,YH:06sh}, most popular models (including stuck-at, bridging and most delay fault models) work on gate level. The stimuli used during testing are usually test patterns either applied through scan infrastructure by automated test equipment (ATE) or generated on-chip by built-in self-test (BIST) or test-compression logic. These patterns are usually produced by automatic test pattern generation (ATPG) procedures that target specific faults. The resulting test set has a fault coverage (number of detected divided by the number of all modeled faults), and there are efficient fault-simulation tools that determine fault coverage.

In contrast, SLT fault models are currently lacking, even though first ideas to define an ``SL-FM'' and to use it to guide ``SL-TG'' to generate PSS scenarios for use in SLT are discussed in \cite{Che:18}. Moreover, the application of conventional fault models, such as stuck-at faults, to SLT is practically infeasible because it would necessitate simulation of very long (billions of clock cycles) SLT sequences for every considered fault.

Regarding the test quality, there is no serious discussion of skipping structural test altogether, replacing it with SLT. Structural tests, in addition to covering defects directly represented by the detected faults, usually also detect a large number of unmodeled defects (``fortuitious detection''), such as manifestations of crosstalk or power-supply noise \cite{WGB:04sh,Pol:10}. Therefore, the ``baseline DPPM level'' is provided by structural tests (it depends on the accuracy of the fault models and the throughness of ATPG). SLT is considered as an additional test insertion if this DPPM level is higher than required by a given application.

The differentiation between \emph{functional test} and SLT is more subtle; in fact, many publications simply treat SLT as a sub-type of functional test \cite{Che:18}. However, we believe that there are serious differences between the ``traditional'' functional test \cite{MHB:00} and SLT. The former is based on running relatively small test programs, usually stored in a microprocessor's cache and written, at least partially, in assembly language. Such programs can be created manually, using evolutionary techniques \cite{SSR+:05sh} or even deterministic test sequence generation \cite{RCS+:16sh}.

Functional test programs can be (and are) assessed with respect to detection of traditional gate-level fault models (stuck-at, delay faults, and the like) or special instruction-set level fault models \cite{TA:80}. As was discussed above, meaningful SLT sequences are far too long to be assessed using these fault models or generated by ATPG. A meaningful coverage metric or any other systematic approach to decide whether an SLT suite is ``good enough'' is currently lacking (see Section \ref{sec:cov} for some ideas to this end).

In many cases, the functional test (which plays also a major role when considering in-field test) is developed targeting single modules in the IC (CPU, peripherals, memories, interconnections). Hence, it basically aims at checking the correct behavior of each single module in an isolated manner. %o On the other side, 
At the same time,
one target of SLT is to check whether the whole device works correctly, exploring for example the effects induced by the interactions among modules. Among the different phenomena triggered by SLT, temperature-related ones play a major role: cases have been reported, where a given defect in an interconnection was only triggered when the temperature gradient between two modules was exceeding a given threshold. Clearly, this kind of defects can hardly be detected by anything different than SLT.

An interesting question is whether functional test and SLT detect the same classes of defects. Both aim at closing the coverage holes of structural tests, and yet one would expect that the expensive SLT would not be applied if the desired test quality were achievable by simpler functional tests alone. One reason for SLT's superiority might be its sheer huge number of test patterns being applied to the circuit during its billions of clock cycles, resulting in a higher chance of fortuitous detections. However, SLT may have systematic advantages, as it is defined on SoC level and incorporates functional interactions between the microprocessor and other SoC components, while functional test tends to focus on the microprocessor itself. For an ultimate answer to this question (and possibly, options to shift some of the detections from SLT to previous insertions), the nature of SLT-unique fails should be better understood. The next section discusses our current understanding of SLT-unique fails.

\subsection{Debug and Diagnosis in SLT Context}
\label{sec:diag}

When doing SLT, determining the root cause of an observed failure is more difficult than for conventional testing, where efficient diagnosis methods \cite{VD:01,HW:09sh} are available. As will be discussed further below, the exact nature of failures observed during SLT is not always clear, and the manifestation of a failure (e.g., a crash) may happen thousands of cycles after its occurrence. Post-silicon validation features such as trace buffers \cite{IGTF:17} or ``quick error detection'' logic \cite{LHL+:14sh} can alleviate this problem. It was proposed in \cite{RHC:19sh} to use machine learning to establish a relationship between SLT failures and values of over one hundred check status registers within a server-grade processor SoC.

The key idea in~\cite{RHC:19sh} is to use a support-vector machine (SVM) to classify SLT failures in the Intel Skylake SoC into one of 18 classes, where each class corresponds to a candidate faulty core or group of cores in the chip. The SVM model was trained using a small number (1,000) of manually created training vectors. A drawback of this approach is that the training data has to be generated manually. Such a manual approach obviously is not practical for high-volume production tests. While it is desirable to use actual fail data from SLT to train the model, a challenge in this context is that SLT fails are ``rare events'', and it would take a considerable amount of time to generate a sufficient amount of fail data  to train the model. An attractive alternative in this context is to leverage transfer learning techniques and methods for root-cause localization methods using streaming data that have been developed for board-level fault diagnosis~\cite{Liu:ITC2019,Liu:VTS2019sh}.

Machine learning-based fault-identification models
exploit knowledge from test results and corresponding
ground-truth data, without requiring a detailed understanding of
the complex functionality of chips. This problem can be
formulated as a \emph{supervised classification} problem, where a
complete set of learning data consisting of pairs $f(x; y)$. Each
instance $x$ is associated with a unique label $y$, and we refer to
each instance-label pair $(x; y)$ as a \emph{sample}. In our application,
since different chips may have different fault candidates,
we train a binary classifier to diagnose each target fault. 
The learning algorithm constructs a classifier that outputs a
class prediction for a given instance.

In a typical SLT scenario, only a limited amount of test fallout data arrives in the early stages of manufacturing. In fact, fallout data and then associated ground truth about root-cause localization arrives in a streaming format characterized by a potentially  large volumes of data instances. As a result, the diagnosis accuracy tends to be low in the early stages of manufacturing. Compared to a static test and diagnosis flow, processing data streams imposes two new requirements on diagnosis algorithms: (1) the ability to adapt to concept drift, and (2) the availability of a limited amount of memory. Online incremental learning algorithms have been proposed for handling streaming data, and  to deal with concept drift, classifiers implement forgetting, adaptation and drift detection mechanisms. To overcome the challenge of limited memory, classifiers record only the key information extracted from the previous round of streaming data instead of all the past samples. Moreover, classifiers can learn the target concepts incrementally instead of training from scratch to save training time. By executing online learning algorithms for streaming data, the trained model predicts more accurately when the data distributions shift. This approach has been utilized for fault diagnosis in printed circuits boards~\cite{Liu:VTS2019sh}. We expect a similar solution to be useful for root-cause localization in SLT.

\begin{figure}
\includegraphics[width=0.98\columnwidth]{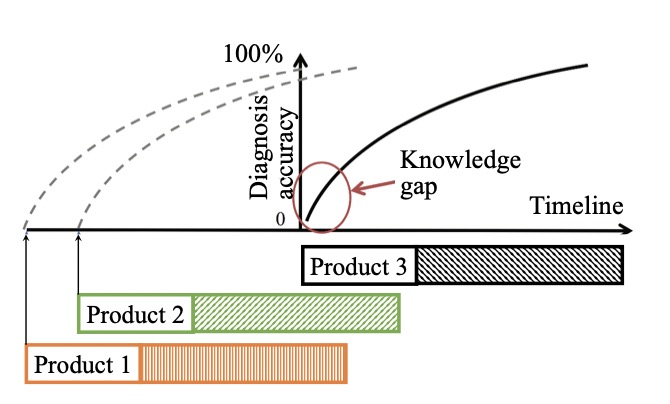}
\vspace{-0.2cm}
\caption{Diagnostic quality over time}
\label{fig:diagqual}
\vspace{-0.3cm}
\end{figure}

The diagnosis accuracy improves when more instances of successful root-cause localization in SLT are available for training the diagnosis system. However, there exists a significant knowledge gap in the initial product ramp-up stage (see Fig.~\ref{fig:diagqual}). In reality, a successful product typically experiences multiple updates and there are often similar products during a period of time. In chip manufacturing, fallout data accumulates over multiple wafers and lots. A similar problem has been addressed for board diagnosis, whereby knowledge learned from a mature board is transferred to the diagnosis model of a new board~\cite{Liu:ITC2019}. A supervised model is trained to identify board-level functional fault using a large number of samples from the mature product (i.e., source domain) and a limited number of samples from the new product (i.e., target domain).

\section{SLT-Unique Fails}
\label{sec:fails}

SLT's main \emph{raison d'\^etre} is its ability to detect failing ICs that are missed during the conventional test insertions. The existing literature suggests a number of sources of these failures, which can be attributed to three broad categories:
\begin{enumerate}
    \item \textbf{Failure mechanisms that are not covered by standard fault models.} For example, traditional scan-based stuck-at and TDF timing tests are node oriented, and explicitly only target the interconnect between the standard cells (gates) in the design. More recently, it was recognized that these classical tests can miss a significant number of defects within complex standard cells, which has led to the development of Cell Aware Tests (CAT) targeting cell internal defects \cite{HHM+:20sh}. Scan tests generated using the CAT methodology can significantly reduce subsequent SLT fallout in some applications. However,  marginal, or ``soft'', timing failures \cite{Sin:19} that manifest themselves only under certain operating conditions (voltage, temperature) \cite{Che:18,BRR+:20sh}, and power-supply instabilities in conjunction with complex power-management schemes \cite{BRR+:20sh}, remain challenging to detect using the available scan-based delay test methods. 
    \item \textbf{Systematic ATPG coverage holes.} Complex SoC designs contain signal lines connecting several, even dozens, of different clock domains. Faults in the logic structures at the clock domain boundaries, asynchronous or analog interfaces or clock distribution networks often need careful manual expert intervention during test generation to ensure reliable detection. These may be conservatively classified as ``untestable'' by automated ATPG tools, even though the faults can manifest themselves during the device's operation. More mundane causes of incomplete test coverage are test time and tester memory limitations. At times some scan patterns that detect only a few faults are dropped to reduce memory requirements or test-application time; this ``long-tail'' problem is also alleviated by fortuitous detection by the very long SLT test sequences.
    \item \textbf{Faults exposed only during system-level interactions.} This includes complex software-controlled clock- and power-domain interactions or resource contention in a multi-core system that cannot be fully replicated on an ATE \cite{Che:18}. Other possible situations include complex hardware-implemented protocols or ``soft'' failures during high-speed memory accesses \cite{BRR+:20sh}. The key characteristics of these failures is that the electrical and timing interaction of the IC under test with other components in the target system  are insufficiently defined, or too complex, to be accurately modelled on a tester.
\end{enumerate}

Since using SLTs as an additional final test screen during post manufacturing tests imposes significant added costs, considerable efforts have focused on minimizing test escapes from traditional structural scan-based testing so as to eliminate, or at least reduce, the need for SLTs. In some applications, SLT is 
%o most critically only needed early during the 
critically needed only during the early
yield ramp phase of production, when the yield of defect free parts is low, as test escapes are obviously more numerous when there are %o many 
more defective parts being tested. Once the manufacturing process matures and yields improve, SLT is only performed in these applications on sample parts for quality assurance purposes---to ensure acceptable defect levels in the shipped ICs. 

An increasing number of high-volume applications, e.g. high-end cell phone processor SoCs, continue to require SLT on all manufactured parts throughout the product lifetime. This has motivated considerable research on understanding the test escapes from traditional scan tests that are uniquely detected by SLTs. The goal is to improve the coverage of these failures by the scan tests, and thereby reduce dependence on SLTs. In revisiting the three categories of test escapes listed above, it is obvious that (3), faults exposed only during system-level interactions, cannot by definition be targeted using traditional test methods. Plugging coverage holes (2) caused by the inability of ATPG tools to generate some tests, e.g., across timing domain boundaries, and for asynchronous and analog mixed signal circuit structures, is a well understood and a longstanding challenge. It is currently the focus of significant development effort at major EDA companies.

However, until recently, considerable mystery has surrounded Category (1), failures that are not covered by fault-model-based scan tests but uniquely detected by SLT. The reason for this uncertainly is the poor diagnostic capability of SLTs. Unlike in scan tests where each applied test input pattern and corresponding circuit response is known, it is extremely difficult to root cause a failure observed by functional SLT down to a logic gate. Failure in functional operation may be observed thousands, even millions, of clock cycles after the underlying logic level malfunction occurs, making it impossible to trace and locate
(see Section \ref{sec:diag}). It is virtually impossible to confirm the cause of most of the SLT failures that escape scan tests. This makes it difficult to target them with new fault models.

The new Cell Aware Test generation methodology aimed at detecting shorts and opens within standard cells has been successful in reducing SLT fallout in some applications, but less so in others \cite{HHM+:20sh}. The distinction appears to be the susceptibility of the design to “soft” timing errors, which are not caused by the “hard” defects targeted by CAT. Particularly vulnerable to timing failures are power constrained applications such as cell phones that implement aggressive power management to ensure battery life while meeting ever increasing computational demands. This involves dynamic voltage-frequency scaling, with the circuit operated at slow clock frequencies and low energy saving voltages when computational loads are minimal. Unfortunately, the impact of manufacturing process variations on circuit timing is greatly amplified during very low voltage operation, causing some circuits to experience occasional timing errors. Ideally, such “defective” ICs containing extremely slow statistical outlier transistors from random process variations should be detected by the scan timing tests. However, the widely used transition delay fault (TDF) model explicitly only targets a single lumped delay in the circuit. It is unable to reliably detect accumulated delays along circuit paths resulting from a distribution of delays across the IC due to the effects of random process variations. What are required are effective scan path delay tests, which have so far not proven practical. Variation-aware tests \cite{SPI+:14sh} are, therefore, not part of the scan test set, resulting in the increasing dependence on SLTs in power constrained low voltage applications.

\section{Assessing the Quality of SLT Programs}
\label{sec:cov}

As was already mentioned above, the standard quality metrics normally used during IC testing (fault coverages) are not practically applicable to SLT, just because the workloads that are being applied for minutes would be impossible to simulate. The only suggestion for an SLT-aware metric proposed so far is the number of used scenarios in the context of the PSS-based SLT flow \cite{Che:18}. In the following, we discuss coverage concepts used during software integration test and whether it can play a role during SLT.

Overall, software integration test is---similarly to SLT---not as
well investigated as unit tests on the lower levels. For the latter,
coverage is often used to describe the completeness of the tests. The
coverage most often encompasses control flow (such as statement
coverage or branch coverage) but can also relate to the data flow
(such as all definitions). This translates badly to integration tests
as it then still relies on which statements in the integrated 
components are executed, just with a relation to the interface
of the components \cite{spillner1995test}.

An alternative approach \cite{HSW:19sh} sees the
integrated components as black boxes. Motivated by the application in the automotive domain, in which many software components are developed by suppliers without revealing the source code, it aims at understanding if the components are well integrated only based on the information from the data flow between them.

To achieve this, 
the observations on shared data between
components are described and classified 
into \emph{preconditions} that represent
the states of the components, \emph{stimulations} that 
capture
manipulations of shared data, and \emph{verification} 
that 
represents
observed data that can be used to check the behaviour based 
together with preconditions and stimulations. This gives rise to new coverage criteria%
, for example, the coverage of all usages of
shared data (``shared-data-use''), or the verification usage of
a shared data (``verification-data-use'') in which a test must
use a shared data to verify behaviour \cite{HSW:19sh}. Using this approach for
automotive integration testing, 
test gaps were identified. 
It appears promising
to use similar 
coverage criteria for SLT.

Some high-level parameters are playing an important role in the detection of failures during the SLT insertion. To stress the component to facilitate the insurgence of an SLT defect is often not a pure duty of the SLT software, but it requires some extra conditions to be met. The characterization of an SLT program must also carefully consider the effects of different temperatures (both high and low) and temperature gradients. 

In most cases when temperature is a factor to consider, the requested temperature conditions are provided by climatic chambers. Anyway, a stressful (therefore valuable) SLT workload should also show self-heating capabilities to reach a specific junction temperature as indicated in mission profiles. Furthermore, it should also implement heat control mechanisms to avoid over stress and sometimes the thermal overrun.

Testing at harsh power supply conditions is another very common industrial practice in SLT; power supply voltage for core and other power domains is increased or decreased even to make the test running out of the functional specifications of the product. This power supply variation is aimed at screening out latent faults that may easily show up in the early life of the IC. Obviously, this may turn into a dangerous practice if the SLT workload was not very carefully graded in term of punctual power demands.

\section{SLT Program Generation}
\label{sec:gen}

While the current state of the art for SLT is to use existing software (thus requiring no explicit generation), testing for specific problems can call for SLT programs with specific characteristics. For instance, if we know that a particular part of the system is vulnerable to subtle failures, an SLT workload that stresses this part of the system will be useful. This is related to the ``power-virus'' generation considered (for much smaller circuits) in the past \cite{NKH+:07sh}. 

Thermal measurement on physical samples like in \cite{BRV+:17sh} is an important practice to also provide physical findings about the implemented SLT workload. Warming up the silicon surface of a device is often asking quite a long time (up to minutes) before the circuit shows the desired (usually high) temperatures. Such a thermal characterization is usually done by thermocamera-based experiments. This experimental observation can be supported by suitable firmware read-out of the chip the temperature recorded by the sensors embedded in it.

The creation of a functional oriented program able to control the thermal behaviour should not address the entire circuit in one shot, but rather focus on different circuit zones at different times. Differently from scan testing, the focused functionalities during SLT should be triggered by a suite of workloads targeting relevant circuit portions.

\section{Conclusion}
\label{sec:concl}

Looking back at conventional integrated circuit test several decades ago, we see a scientific success story, which made quality assurance feasible and kept its cost reasonable throughout the long period of exponential circuit complexity growth. A (largely) common understanding of terms and concepts throughout a large community, both industrial and academic, has led to sophisticated and yet practical solutions that could be adopted by most relevant players. We believe that SLT needs to reach a similar level of widely agreed-upon understanding  to enable comparable progress and overcome its foreseeable limitations. To answer the questions posed in this paper, a new level of cross-sector collaboration between semiconductor manufacturers, system integrators, test equipment manufacturers, EDA companies and academia will be necessary. For instance, reliable information on SLT programs that are practically effective, and on SLT-unique fails identified by such programs, will be helpful in giving the necessary research, e.g., on improved generation of SLT programs and systematic assessment of their quality, a meaningful direction.

\section*{Acknowledgment} 

This work was supported by Advantest as part of the Graduate School ``Intelligent Methods for Test
and Reliability'' (GS-IMTR) at the University of Stuttgart.



\bibliographystyle{IEEEtran}
\bibliography{references-slt}

\end{document}